# QUANTUM/RELATIVISTIC COMPUTATION OF SECURITY AND EFFICIENCY OF ELECTRICAL POWER SYSTEM FOR A DAY-AHEAD

Stefan Z. Stefanov

ESO EAD, 5, Veslets Str., 1040 Sofia, Bulgaria

szstefanov@ndc.bg

Abstract. An algorithm for Electrical Power System (EPS) quantum/relativistic security and efficiency computation via perturbative renormalization of the EPS, finding the computation flowcharts, verification and validation is built in this paper. EPS renormalization is performed via virtual thermalization. EPS energy renormalization provides EPS critical load and outlines the EPS fractal landscape. EPS time renormalization gives the critical exchanges in EPS and shows the electricity minimal leveled cost and the market price of the electricity. The computation flowchart is obtained through market homogenization, virtual market shocks, Random Matrix Theory-communication at the market and electricity flow in EPS. This flowchart causes entropy stochastic resonance in EPS and performs EPS connection percolations. Computation verification is achieved via checking the EPS separability and checking EPS grid. Computation verification consists of checking the solving problems for security and efficiency and checking the changing in EPS, as well in checking of the forecasting for EPS. Computation validation is carried out through forecasting and avoiding a threat for EPS. The partial valid computation is recorded in Programming language for Computable Functions-style and causes limited rational action of the coarsed EPS.

**INTRODUCTION**

Prediction and planning is the highest Smart Grid intelligence level (NEMA, 2009). The EPS automatic analysis is performed on this level in order to enhance the system leading. This includes any system-wide application of advanced control technologies, such as devices and algorithms that will analyze, diagnose, and predict conditions and take appropriate corrective actions to eliminate, mitigate, and prevent outages and power quality disturbances. Exterior factors, such as the current and potential state of the EPS environment, could be factors at these new technologies. Resource management, timing, and using the external variables are characteristics of (NEMA, 2009) prediction and planning in Smart Grid.

The central unit, which predicts and plans in the Smart Grid, is a monitor for the EPS probabilistic reliability (Sobajic, 2003) as well a watch (Sobajic and Douglas, 2004).

Assessment of this unit is upon the criteria for the Smart Grid assessment by (Scott, 2009). This is assessment for the predicting and planning complexity.

The objective of quantum/relativistic computation of the EPS security and efficiency is to build 'Daily Artificial Dispatcher' (DAD), i.e. a Smart Grid central unit that predicts and plans for a day-ahead.

Quantum/relativistic computation of security and efficiency of EPS will be done as regularized computation by (Manin, 2009a) and (Manin, 2009b). This means that the security and efficiency of the power system will be sought by perturbative renormalization of the EPS for a day ahead. Because of that, DAD will be built as harmonic composition of predicting for day-ahead programs. Here the predicting for a day-ahead program is a description of method for calculating the predicting for a day-ahead function, according to (Manin, 2009a) and (Manin, 2009b). This harmonic composition is a stable structure of synchronized, predicting day-ahead programs.

DAD will be searched for as a post modern fairy tale by (Lyotard, 1993), which to organize these predicting for a day ahead programs. This post modern fairy tale generates (Losh, 2007; Moslehi, 2010) Smart Grid.

DAD, obtained by regularized computation and presented as a postmodern fairy tale, is according to (Riedl and Young, 2006), a hermeneutic network by (Zhu and Harrell, 2009). Then DAD solves the problem for security and efficiency of the EPS for a day ahead in terms of Pascual-Leone. Indeed, the hermeneutic network is the solution of the problem in terms of Pascual-Leone, according to (Shannon, 2008). The reliability and capacity of this hermeneutic network determine the security and efficiency of the EPS for a day ahead. This security and this efficiency are recognized as a decoupled fixed point of the regularized computation, according to (Manin, 2009a, b) and (Jerome, 2002).

DAD, as a hermeneutic network, resembles the 'dissipative' brain of (Freeman, 2009) and (Vitiello, 2009). Therefore, DAD will be investigated as quantum/relativistic computation via gedanken experiment.

DAD is presented in (Stefanov, 2010). In this work is investigated DAD realization as a quantum/relativistic computation through EPS perturbative renormalization, generating the computation flowcharts, verification and validation.

In Chapter I of this paper have been carried out the perturbative renormalization of EPS energy and time for a day ahead via virtual thermalization of the EPS for a day-ahead (Stefanov, 2007).

In Chapter II of this paper are done the very EPS quantum/relativistic computation through predictive analysis, according to (Stefanov, 1992, 1994), of the EPS and market synchronization and steady state reaching. So, in this way, the EPS quantum/ relativistic computation flowchart is obtained.

In Chapter III of this paper is tested EPS quantum/relativistic computation by separability and network of (Afraimovich and Glebsky, 2003).

In Chaper IV of this paper EPS quantum-relativistic computation is validated through EPS logic under threat. In this case EPS state under threat is considered as squeezed coherent state by (Vitiello, 2009).

# Chapter I. RENORMALIZATION

## 1. EPS LOAD

ENERGY RENORMALIZATION 1. Virtual thermalization of the 'Elastic/Plastic' model of the 'EPS – Market' system provides the EPS critical load.

### 1.1. EPS LOAD SHOCK CHANGE

Critical change of the EPS daily load is the result from linear and nonlinear operation with boundary power of the thermal weather machine having cold and hot heat source. The boundary power and entropy production of the thermal machine with cold and hot source, in linear and nonlinear operating mode, are found according to (Tsirlin, 1997).

This critical change of the load and this critical change in entropy are observed holographically (Hartnoll, 2011) as surface water waves.

Total critical change in the load (Bruno, 1998) is obtained by the surface water waves.

Calibration of this general critical change is done through the results for the average load change by (Rahman, 1990) and (Bunn and Farmer, 1985).

This total critical change develops at the speed of shock water wave (Comets, 1991) and travels, per unit of time, the distance between the atoms of a molecule (Domenicano and Hargittai, 1992). This molecule is observed at the holographic screen.

The distance, traveled per unit of time for the total critical change, is the path of the open thermodynamic system, corresponding to EPS (Stefanov, 2006).

### 1.2. EPS LOAD SYMMETRICAL CHANGE

#### 1.2.1. OSCILLATORS RING

Energy rotation in EPS is a degenerative mode in the ring of three Van der Pol oscillators (Ookawara and Endo, 1998). These three oscillators are defined by angular frequencies of the EPS model by (Stefanov, 2001) and (Stefanov, 2003).

The oscillators' inductance is derived from the synchronization in the ring. The oscillator resistances are determined as the weights of the expected loads of Regional Dispatching Centers, in regard to the total expected EPS load. Resistance and inductance calibration is done according to the results by (Zhong et al., 1998) and by (Ookawara and Endo, 1998).

The expected mode length in the ring is calculated as a path by (Stefanov, 2006).  The expected droop is determined from the results in the same paper.

### 1.2.2. SPECIFYING THE EPS LOAD

Specifying the expected loads in EPS and the EPS expected angular frequencies is modeled as a Hamiltonian invariance of the system 'EPS-Market' for symmetries at transmission and for symmetries of the grid.

The transmission symmetries are symmetries when change in the weather takes place on quadratic nonlinear circuit by (Armenskiy, 1989). Symmetries in transmission are found according to (Bachmann and Schmidt, 1970). Weather change is: a / change from the northwest b /  change from north or from the west. Circuit is with generated voltages that are the expected loads of Regional Dispatching Centers, and with impedances defined above.

Symmetries of the grid are symmetries in the impedance ring.

Both types of symmetries from here define quantum encoding according to (Kish, 2009).

Critical change of the daily EPS load from § 1.1 is defined on a triangular grid. The load specification here is defined on the square grid.

### 2. EPS EXCHANGES

ENERGY RENORMALIZATION 2. Virtual thermalization of the 'Elastic/Plastic' model of the 'EPS – Market' system outlines the EPS fractal landscape.

EPS exchanges are observed holographically (Hartnoll, 2011) as auto potential anomaly of a horizontal cylinder, buried in the ground. Exchanges are observed as such potential after (Levitin et al., 2005). This self-potential as well the location of the horizontal cylinder underground, are found according to (Babu and Rao, 1998).

### 2.1. SELF-POTENTIAL EPS RECONSTRUCTION

The exchanges self-potential is reconstructed in relation to the weather evolution sound. The weather evolution is given through Lyapunov exponent as stated by (Prigogine, 1980). The sound of this evolution is defined as the sound of heat carrier movement in the pipeline with a rectangular profile by (Baranov et al., 1998). Holographic observer from (Hartnoll, 2011) sets this pipeline.

The pipeline potential is reconstructed by the heat carrier turbulence.

Carried out reconstructions provide the heat exchange between the atmosphere and the water surface. This heat exchange is heating water, according to (Levitin et al., 2005). The heat exchanger for water heating is calculated according to (Kaya and Chen, 1980). Its energy efficiency is found after (Voigt, 1978).

$T^{\frac{1}{2}}$ -thermalization from §1 is added here with T –thermalization.

## 2.2. EXCHANGES POTENTIAL CORRECTION

EPS is considered as composed of two subsystems, namely a subsystem with active power exchange and a subsystem with reactive power exchange. Stability of this EPS depends on the EPS droop. The droop $k_c$ defines the critical EPS trajectory deviation from a single sphere. This critical deviation can be found by the results for stability of multivariable system by (Borisenco et al., 1988) and from information entropy of each of the subsystems (Saridis, 1984), (Bellman, 1960).

This critical deviation is due to the impulse duration $k_c$, emitted by the disk, describing the rectangular area of the holographic screen. This rectangular area is with dimensions defined by the first cummulant of the exchanges matrix trace. The results for the first cummulant of the noncentral Wishart matrix from (Mathai, 1980) are used here.

This critical deviation is considered as a field wavelet. This critical deviation is due to the impulse of the former type, according to the solution of the inverse source problem for wavelet fields (Devaney et al., 2008).

Self-potential of the exchanges is corrected by the impulse generated by the droop.

## 2.3. EPS FRACTAL LANDSCAPE

Fractal landscape is self-similar fractal bubble, according to (Winitzki, 2005). Spectral dimension of a fractal landscape that is 3d-sector is $D_s = 1.5$, according to (Modesto, 2008). Spectral dimension of the fractal landscape that is 4d space-time is $D_s = 2$, according to (Modesto, 2008).

EPS, as self-potential of a buried cylinder, is isospectral to the fractal landscape, according to (Levitin et al., 2005), (Band et al., 2008). Here, fractal landscape and EPS are considered as isospectral surfaces with Dirichlet-Neumann mixed boundary conditions.

Dual area of EPS-area in terms of elasticity and plasticity (Sadovskiy, 1997) is a spherical cap, which generant is a shock wave.

The EPS potential is synchronized with the potential of the shock wave. Beating between the two potentials gives the deviation from EPS synchronous time. This deviation is obtained from the results in (Hamilton et al., 2007).

EPS fractal dimension of a fractal landscape, which is 3d-sector, is the fractal dimension of 2d-curved space on multiple scales. Fractal dimension of 2d-curved space on multiple scales is obtained from (Barth et al., 1993). It is a photography of time $T_{16}$ as a string with a length L=32. Here $T_{16}$= 2L, according to (Dáger and Zuazua, 2002).

EPS fractal dimension for a fractal landscape, which is 4d-space-time, is Hausdorff dimension of the fractal landscape surface. This is photography of the deviation from EPS synchronous time.

EPS–area and spherical cap are visualized in the screen plane. Deviation from the synchronous time is shown on the screen as a deviation from the diamond center, which represents the spherical cap. EPS is visualized as a serpent as well as spectral bands. Here 2d-distorted scalable space is represented as a serpent, and the fractal surface - as spectral bands.

## 3. ELECTRICITY COST/PRICE

TIME RENORMALIZATION 1. Virtual thermalization of the 'Energy/Cost' model of the 'EPS – Market' system gives the minimal electricity leveled cost and the electricity market price.

### 3.1. MINIMAL ELECTRICITY LEVELED COST

Let a dispatcher compensates 90% of coming load change. Dispatcher's benefit is given by the obtained compensation. Here the compensation of the load change is according to (Ho et al., 1985).

The load is deformed by 10% uncompensated change. The deformed load defines a new EPS droop. This droop $k_c$ provides the following dispatcher's benefit B, B=1 – ln $(k_c)$/2. The dispatcher's benefit is this, according to (Alimova and Trushkov, 2008), when dispatcher leads optimal EPS at constant exchanges.

Let the dispatcher exchanges as much energy as is the energy of uncompensated load. The benefit from these exchanges is obtained from the wind benefit and from the heat engine benefit, because the weather turbulence defines the uncompensated load. Here the wind and the heat machine are regarded as fuzzy energies according to (Monteiro and Miranda, 1997).

The wind benefit and the heat machine benefit are modeled with a triangular distribution of the dispatcher's benefit B. Here the triangular distribution is as in (Crovelli and Balay, 1991). These two kinds of benefits differ fuzzy. Here the fuzzy distinction is a distinction of the triangular forms by (Girouard and Polterovich, 2008). The fuzzy distinction presents the exchanges benefit as a weighted sum of the wind benefit and the heat machine benefit. The electricity leveled cost equals this exchanges benefit.

The electricity leveled cost is minimal when the exchanges benefit is maximal. This minimal cost is obtained via deformation of the peaks and valleys of the expected load.

### 3.2. ELECTRICITY MARKET

Expected home market is presented as electricity auction. This market is calculated in (Stefanov, 2003).

The expected intersystem electricity market is analogous to the electricity wheelling in Regional Dispatching Center (RDC) – model of the system 'EPS-Market'. RDC model is a compartment model with three compartments. The three compartments are three RDC, considered as homogeneous systems. Wheelling is exchanges in circle among the three RDC of the compartment model without input and output (M-model). This M-model is transformed into a compartment model with input and output (J-model) according to (Godfrey and Chapman, 1990). Intersystem exchanges are the difference between the input and output of the J-model. Because M-model and J-model are indistinguishable, according to Godfrey and Chapman (Godfrey and Chapman, 1990), the upper expected intersystem electricity market is truly modeled.

The expected price of the exchanges for such an intersystem market is given by the exchanged amounts of energy.

### 4. FLOW IN EPS

TIME RENORMALIZATION 2. Virtual thermalization of the 'Energy/Cost' model of the 'EPS – Market' system gives the critical exchanges in EPS.

Expected EPS exchanges are turbulent at virtual EPS thermalization. Heat exchange in this virtual thermalization is recovered via Kalman filter from (Melsa and Jones, 1973).

The expected EPS exchanges are both turbulent and dependent on the serpent of the electricity cost. The mechanical realization of the open EPS by (Stefanov, 2003) gives this second relationship.

The above expected exchanges, when EPS is considered as a system with stationary source, are found as a linear solution of type 'localized solitary traveling wave' of the Burgers' equation for a stationary source (Petrovski, 1999).

The above expected exchanges, when regarding EPS as non-commutative field or as a system with a non-stationary source, are found from the EPS potential.

It is assumed that EPS is with two related states - the ground state and the excited state. It is assumed also that Hamiltonian of this EPS is pseudo-Hermitian Hamiltonian (Ahmed, 2003). EPS states are found by Padé approximation, obtained via variational principle (Baker and Graves-Morris, 1981), of the EPS wave functions.

In examining the EPS as a non-commutative field, the EPS exchanges are shock soliton, which is formed by two exponents. This soliton is found as a solution of the Burgers' equation for non-commutative space-time (Martina and Pashaev, 2003).

In examining the EPS as a system with non-stationary source, the exchanges are auto-model solution of Burgers' equation with non-stationary source (Petrovskiy, 1999).

Time reversal in EPS gives the flow in EPS. At first the EPS Hamiltonian's spectrum is found according to Rosu (Rosu, 2003). The flow in the EPS is found from a Parrondo's game for the EPS Hamiltonian's spectrum. Such a game is generated according to Toral (Toral et al., 2003) by the time reversal in EPS.

The above expected exchanges, in considering EPS as a non-commutative field, are obtained from the flow in EPS. It is assumed that EPS is with a gradient flow and with the diffusion semi-group in metric space. Optimal flow in this EPS is found according to (Savaré, 2007) and (Stefanov, 2003). This consideration of EPS leads to the exchanges, which are shock soliton formed by an exponent. This soliton is found as a solution of Burgers' equation for non-commutative space-time (Martina and Pashaev, 2003)

## Chapter  II.  FLOWCHART

1.  SYNCHRONIZED ELECTRICITY MARKET

FLOWCHART 1. Random Matrix Theory (RMT) - communication of market participants synchronizes the electricity market with EPS.

1.1. EXPECTED POWER TRANSPORT

Let's consider the system 'EPS-Market' as a disordered system with transport of energy. The complexity of this disordered system is given by transport. Here the transport in disordered system is understood according to (Kenkre et al., 2008) as transport in medium where the spatial disorder is replaced by temporal memories. These temporal memories are assigned by a diffusion market, defined according to (Dokuchaev, 2001).

The spatial disorder is presented as customers' size distribution, as a deviation distribution from the synchronous time and as distributed prices.

Customers' size distribution is found from the expected daily EPS load. This is obtained as a typical log-grain size according to the results from (Christiansen and Hartmann, 1988) and (Artyushenko and Nikitenko, 1998).

Deviation distribution of the synchronous time is obtained from the expected EPS frequencies. These frequencies are found in Chapter I. They determine the deviation distribution from the synchronous time when time is modeled as a random variable with a stable distribution. This deviation is derived from the results of a stable distribution by (Fofack and Nolan, 1999).

Distributed prices are determined as generation and consumption prices of the competitive market between them and the Regional Dispatching Center (RDC) prices at closer look at EPS.

Winning strategies at the diffusion power market are approximated according to (Mainardi et al., 2000).

Electricity transport shows the confidence in Transmission System Operator (TSO) - dispatchers and the likelihood of electricity prices. Electricity transport reveals the complexity of the system 'EPS-Market' as confidence and likelihood. This disclosure is done by the results of (Afraimovich and Zaslavsky, 2003) for space-time complexity of Hamiltonian dynamics. This system 'EPS-Market' is adopted for RMT-medium and hyperbolic dynamic system by (Collet and Eckmann, 2003).

Extreme price of electricity is calculated according to Vasicek (Weron et al., 2001).  In addition, in this calculation are reported 'EPS-Market' fractional medium, medium spectra and energy distribution in the medium. The border of medium spectra is obtained according to (Lehner, 2000), when the medium is modeled as a free sum of EPS and market. Free probability of medium energy distribution is determined according to (Bercovici and Pata, 1999).

The expected load change is found as the 'EPS-Market' system perturbation. This change is obtained from the (Vittot, 2003) results for management with small perturbation of the chaotic transport in Hamiltonian system, where the system 'EPS-Market' is considered as Hamiltonian system with chaotic diffusion.

Optimal distribution of the purchased electricity of the diffusion market determines the optimal entanglement at this market diffusion, and hence - the risk of the considered diffusion market. This risk is obtained according to the results for a purchase distribution from (Liu and Guan, 2003).

## 1.2. EXPECTED CURRENT PRICE

Let the expected electricity current price is a price of an option with fractional noise from (Vilela Mendes and Oliveira, 2004). Then the expected current price is found as one-parameter isospectral function of the expected energy of the system 'EPS-Market'. This is done by fractional transformation of the Riccati equation of constant coefficients from (Rosu et al., 2001) and (Reyes et al., 2003). Here the randomness at a macroscopic level is considered as fractionality.

The expected flow determines another estimated current price. This is done by heat kernel expansion on the integers (2, 3) and (3, 2) from (Grünbaum and Iliev, 2003).

## 1.3. EXPECTED PRICE FLUCTUATION

Let the expected price fluctuation of electrical power is caused by a rare random event, as this happens in (Sánchez et al., 2003). This expected price fluctuation is a spectral fluctuation of RMT- medium by (Ahmed and Jain, 2002). Then the expected power transport is a separatrix-like motion by (Iomin et al., 2002) in this RMT- medium. The expected electricity transport time is found from the expected price fluctuation according to (Cherniha and Henkel, 2004).

## 1.4. PROFIT DUE TO THE EXPECTED TRANSPORT

Let the generation and consumption in the system 'EPS-Market' are entangled according to (Rosu et al., 2004) with a Dirac-like coupling parameter. Then the expected transport time is found according to (Rosu et al., 2004).

The conflict between generation and consumption is a conflict of normal and anomalous transport by (Weitzner and Zaslavsky, 2002). This conflict is a pursuit game by (Chikrii and Eidelman, 2001). The transport time in such a conflict is found from the terminal set of the above game according to (Chikrii and Eidelman, 2001).

The TSO profit is the product of the transport fluctuations and price fluctuations of the electrical power. Here the profit is found for the electricity market that is a financial steam engine  according to (Khrennikov, 2004). Transport fluctuations are found according to (Gukov, 2003) for electricity sale, which is at a crossroad.

## 2.  STEADY STATE ELECTRICITY MARKET

FLOWCHART 2. Homogenization of the market leads to a steady state market.

The similar behaviors at the electricity market are recognized by the market patterns. This similarity is the dimensionless moment of the market waves. This dimensionless moment is found for the perturbed market oscillations, referred to the 2-tolerant development and 2-tolerant market cyclic behavior. This moment is obtained according to the results for pattern reconstruction through dimensional analysis by (Melan and Rudolph, 1999). The perturbed market dimension is found by (Vittot, 2003).

Wave, development, and cyclic behavior of the market are found by the irreversibility of the market behavior. This irreversibility is considered as a breaking of symmetry between the market cyclic behavior and market development. Breaking of this symmetry is related to the oscillations of two damped/amplified parametric oscillators in accordance with (Alfinito and Vitiello, 2002). Wave, cyclic behavior, development and thermodynamics of the market are derived from the results of Alfinito and Vitiello for these oscillations.

The electricity market is a cooperative game of 4 +1 players. In this game, TSO is trying to sell electrical power to all four agents or to some agents only. Here the four agents are identified with the four cardinal points. Thus the nominated electricity market is a game for informational trade. This game is given by the energies of the four market oscillations. These energies are determined by the relationship between oscillations and distances by (Binder, 2002) and by natural neighbor interpolation by (Bern and Eppstein, 2002).

An optimal energy distribution in this cooperative game is found as the Shapley value.  This value is found according the regularized Shapley principle by (Petrosian and Kuzyutin, 2000).   The so found Shapley value is dynamically stable for deviations from the synchronous time. These variations are modeled as flicker noise according to (Planat, 2004).

The optimal Shapley value is the above game value or the value of a subgame of three. This optimal value shows the similarity in energy of the electricity market players.

The proximity of the above electricity market, to the classical one, provides EPS  security.  This  proximity  is  calculated  as  the  distance  between quantum/mechanical system and a classical mechanical system, according to (Abernethy and Klauder, 2004). Also, this proximity is calculated as a rotation of the above market in comparison to classical one, according to (Breuer, 2003). The EPS security, found in this way, is a quantum Boolean function, represented by its Fourier coefficients according to the results for quantum Boolean function by (Montenaro and Osborne, 2008).

Waves, observed at electricity market, homogenize the market. This homogenization is done as a restoration according to (Barlow et al., 1995) of the macroscopic isotropy of fractal medium with microscopic anisotropy. Medium conductance and fractal medium transport are determined, respectively, by the superimposition of the similarity in energy and the similarity in behavior and by the entanglement of these two similarities.

Temporal planning of the electricity market is through the dispersion of trading sessions. This temporal planning is understood as in (Takács et al., 2003). This dispersion is found when examining the trading sessions as the separatrix splitting near multiple resonances, and evaluation of the corresponding stochastic process. Then, the dispersion is obtained from the results of separatrix splitting near multiple resonance, of (Rudnev and Ten, 2005), (Quesne et al., 2003), (Kolodyazhny and Rvachov, 2004) and from the results for estimation of Lévy processes by (Figueroa-López and Houdré, 2004).

Cyclic planning of the electricity market is done by aligning the actions of four agents in the upper cooperative game among them. The error of this planning is given by the inability to align the actions of the four agents. This error is found from the results by (Banderier et al., 2004).

Specifying the market planning is achieved by the similarity in energy and the similarity in the market behavior. Specified dispersion of that market is found when examining the trade as a stochastic process for homogeneous random fields. This specified dispersion is obtained from the results for homogeneous random fields by (Malyarenko, 2004), (Pospisil, 2004) and estimation results of Lévy processes (Figueroa-López and Houdré, 2004).

The EPS droop is obtained from the market turbulence or from the market potential. Here the market turbulence is found from the results of turbulence by (Eyink and Wray, 1998).

The EPS security for the snake of electricity market prices ( $p_r$ ) is defined from the market homogenization when it provides a finite number of opportunities. This security and homogenization time of this market are found from the results for a Sierpiński gasket by (Barlow et al., 1995).

EPS security for the electricity market spikes (quaking the market) ($p_v$) is defined from the market homogenization when it provides unlimited opportunities. This security and the homogenization time of this market are found from the results for Sierpiński carpet by (Barlow et al., 1997) and the results from quaking by (Sibani, 2005).

## 3. SYNCHRONIZED EPS

FLOWCHART 3. Electricity market surprises synchronize EPS with the market.

### 3.1. RISK IN 'EPS-MARKET' SYSTEM

The risk in the 'EPS-Market' system provides the electricity market surprises.

The risk in the EPS-'Market system is a risk of intentions for this system. Then the risk in the 'EPS-market' system is due to unfocused or uncertain electricity bid, according to the understanding of intent by (Tan, http://ezinearticles.com ).

The electricity bid is unfocused when there is an energy collapse or when there is anomalous diffusion of the generation and consumption. This energy collapse is considered according to (Fibich, 2009), and the anomalous diffusion of generation and consumption is anomalous diffusion in fusion plasma (Ciraolo et al., 2003).

The electricity bid is uncertain, when there are energy fluctuations or when there is a profit singularity. Here the energy fluctuations are fluctuations by (Bussi and Parrinello, 2008), and the profit singularity is by (Davydov and Mena-Matos, 2005), (Barone-Adesi and Gigli, 2003).

The surprise of the unfocused bid is heard. It is heard as hearing the fastening of a rod by (Akhtyamov et al., 2008).

The surprise of uncertain bid can be seen. It is seen as an interesting event is seen when watching video by (Itti and Baldi, 2005).

### 3.2. MARKET POTENTIAL

Let the electricity market is quantum integrable system, driven by a hidden force. Energy spreading and the potential are expressed by symmetric functions of (Baseilhac, 2006). Energy spreading is found at sudden market change according to (Hiller et al., 2005). Market potential is found for the expected market rip according to (Andrianov et al., 2005).

### 3.3. DYNAMICAL MARKET STATE

The market is modeled as a replicating two-strand system dynamics. It is assumed that the market has two states (normal and excited), and that the two strands are leading and lagging strands in opposite directions. This market is modeled by (Aerts and Czachor, 2005).

States at this market are energy normed through the results by (Erlebacher and Sobczyk, 2004).

Also, the market is modeled as a linear Hamiltonian system with double-resonance. This is a dual model of the strands model. States in this market are found from (Markeev, 2005).

These two market models comply with two modes of reaction-diffusion in the `Prisoner's Dilemma` game. These two modes are obtained by (Ahmed and Hassan, 2000). The first of these modes determines nonlinear wave, which is a string. The existence time and the energy of this string are connected (Joung et al., 2006) as

$$\sin(E^{\frac{1}{2}}) = \pi \sinh(t/10)$$

The second mode determines nonlinear wave, obtained according to Markeev.

The accurate estimation of market state is derived from entanglement of the wave market model with the strand market model and with the resonance market model. This entanglement happens according to (Zalar and Mencinger, 2003).

## 3.4. MARKET ENERGY AND NON-EQUILIBRIUM

The market evolution is determined by the market energy and market non-equilibrium. Here market non-equilibrium is found by (Başkal and Kim, 2006)

The market evolution is presented as statistical submanifold evolution surface, using the reversible entropic dynamics of (Cafaro et al., 2007).

Market development times are found when examining the market non-equilibrium as a parity violation and as arrow of time. These development times are obtained from (Asadov and Kechkin, 2006). The market evolution is indeed such a development. The test for this is done according to (Darné, 2003).

Market evolution changes by the economic medium turbulence. Turbulence changes the energy of the snake motion of the electricity price. This change is found as flexible structure motion change according to (Stam, 1997). The turbulent wind changes the energy of the hidden force that drives the market. This change is found as a Jordan rigid body motion change according to (Crasmareanu, 2002), (Imbert et al., 2000). The standard turbulence deviation is obtained from the relative hydrodynamic permeability and the relative electrical permittivity of the economic medium. These two economic medium characteristics are calculated from the fluctuating market state and from the time fluctuations of market development. The economic medium is assumed as non-equilibrium thermodynamic medium with electrokinetic phenomena (Ageev, 2005) in it.

The turbulence work for market evolution change is a work of a quantum Carnot heat engine, based on a harmonic oscillator. This work is obtained by (Quan, 2006). Minimum turbulence work determines the electrostatic market potential.

3.5. SYNCHRONOUS TIME DEVIATION

The expected electricity market induces expected synchronous time deviation.

The expected synchronous time deviation is determined from the grid correlations, caused by the market. These grid correlations correspond to electronic correlations in the solid by (Eckstein, 2007).

The electronic correlations lead to difference between the chemical potential and the electronic interaction. Their corresponding grid correlations are obtained as differences of energies/prices, when the energies/prices are related by energy-area inequality of (Morgan, 2004).

The energy differences are two pairs. The first two differences are:

1/ Difference of the market energy and the snake price energy when this snake is represented as a fractional order intermediate wave according to (Engheta, 1999);

2/ Difference of the virtual EPS energy and the energy of the EPS electric field, when this field is represented as harmonic map according to (Morgan, 2004).

The second two differences are:

1/ Difference of the snake energy and the market doubled heat rejection surface;

2/ Difference of the EPS electric field energy and the doubled splitting area of the separatrix 'C' (Consumption) and the separatrix 'G' (Generation) of the market according to (Treschev, 1996).

Price differences are also two pairs. The first one is the difference between the expected price for C/G and the predicting market price for C/G. The second one, from these pairs, is the difference between the predicting market price for C/G and the expected deformed prices for C/G, obtained by deformation according to (Teplinsky, 2007).

3.6. CASCADING FAILURE IN EPS

Cascading failure by (Carreras, 2009) is considered as exponential expansion by (Ahmed and Rideout, 2009). Then the critical limits of the expected deviation from the synchronous time are determined by the transport time $T_{16}$ and by the transition from transport to lack of transport. Here the transition from transport to lack of transport is presented as a metal-insulator transition by (Eckstein, 2007).

The transition from transport to lack of transport is found by the above differences and the market temperature. This transition is specified by the EPS magnetism. The EPS magnetism is found according to (Castelnovo, 2009) from the expected EPS reliability, with respect to a rare event and concurrent phenomenon and from expected reliability, with respect to a cooperative phenomenon.

The expected EPS viscosity is determined by the quantum walk for the expected time deviation from the synchronous time. EPS here, is considered as a grid of type 'K$_{3,3}$' and the quantum walk is calculated according to (Jafarizadeh and Sufiani, 2007).

4. EPS STEADY STATE

FLOWCHART 4. Electricity flow, determined by the market, sets EPS in a steady state.

4.1. EPS PREDICTIVE ANALYSIS

EPS security and effectiveness predictive analysis is made by thermalization (Stefanov, 2007). Thermalization induces a flow in EPS.

The EPS behavior, determined by this flow, is obtained when considering EPS as a completely integrable system with unconfined singularities and with Ricci flow in it. Times and exchanges, that characterize this EPS behavior, are obtained from the results in (Mañosa, 2007; Di Cerbo, 2007). EPS droop, at this behavior, is obtained according to (Ponno and Bambusi, 2004). The electricity market, at such a flow in EPS, is presented as a quasi periodic system by (de Souza et al., 2007).

4.2. FLOW IN EPS

The flow in EPS deforms EPS behavior. This deformation is obtained as a difference of the EPS Hamiltonian and the electricity market Hamiltonian. The EPS Hamiltonian is obtained here as Jordan rigid body Hamiltonian, according to (Crasmareanu, 2002). The electricity market Hamiltonian is found for a market that is a quantum system with gauge/string duality according to (Klebanov, 2006). Trigonometric polynomial flow in EPS is found according to (Quispel, 2003) from EPS deformation.

The EPS rate of change, induced by the flow in EPS, is the rate of change of the classical integrable field when there is a defect in it. This EPS rate of change is obtained from the results for a forced field by (Pasquero, 1999) and (Caudrelier, 2007).

Flow in EPS causes oscillations in EPS. The average frequency of these oscillations and the phase frequency of these oscillations are obtained from the EPS droop according to (Rossberg et al., 2004).

EPS self-healing is monochromatic tightening in the knot on a torus. The number of crossings of this knot is found according to (Willerton, 2002) from the electricity market Hamiltonian. The new EPS transport time is obtained as upper bound for ropelength on EPS torus. This length is calculated by the number of crossings of this knot, according to (Cantarella et al., 2003). The energy of the EPS tightening on torus is found from the universal growth law for knot energy by (Lin and Yang, 2008). According to this law, the energy of the EPS tightening on torus is equal to the number of crossings of its corresponding knot at degree 0.92.

4.3. EPS MAGNETIC POTENTIAL

EPS magnetic potential specifies the angular frequencies and probabilities of failure of the EPS.

Oscillations frequency square of one-dimensional harmonic oscillator is equal to the Hamiltonian oscillator fluctuation (Bussi and Parrinello, 2008). On the other hand, this fluctuation is equal to the probability square of oscillator failure. Then, EPS oscillations caused by a failure in it are three. These oscillations are EPS oscillation as an electric field oscillation, EPS oscillation as a magnetic field oscillation and EPS oscillation, forced by the flow with frequency set by the product of EPS droop and the flow. These oscillations determine EPS motion as Langevin dynamics of a driven charged particle in a magnetic field. This EPS motion is obtained from the results of (Jayannavar and Sahoo, 2007) for such a dynamics.

EPS magnetic potential is determined by the work of this EPS motion. The specified angular frequencies and the probabilities for EPS failure are obtained by this magnetic potential.

4.4. DEVIATION FROM SYNCHRONOUS TIME

Hypothetical EPS thermalization causes gauge correlation. For this gauge correlation there is a corresponding financial correlation in portfolio optimization. Therefore, the expected deviation from the synchronous time is found from the (Hamilton et al., 2007) results for gauge correlation and (Pafka et al., 2004) results for financial correlation in portfolio optimization.

4.5. EPS DROOP

The expected EPS magnetic potential gives the expected EPS droop. The EPS oscillations, caused by the EPS magnetic potential, are considered as topological singularities and the transformation with magnetic potential – as a transport in kicked Harper model. The droop is found from the Satija graphics (Satija, 2005) for kicked Harper model.

5. FLOWCHART'S ACTION

FLOWCHART'S ACTION. The flowchart causes EPS bond percolation and entropic stochastic resonance in the system 'EPS-Market'.

### 5.1. EPS BOND PERCOLATION

The expected EPS bond percolation provides the market for the expected system 'EPS-Market'. EPS transformation with bond percolation determines the correction of EPS ($k_c$, $dt_s$) and a new EPS state after this correction. EPS adiabatic transformation energy through bond percolation is found by the results of (Loh et al., 2007).

This transformation consists of three transformation pairs:

1/ 'Triangle-star' (Zhang et al., 2005) and 'star–triangle' (Bibikov and Prokhorov, 2008);

2/ 'Spectral gap' (Verstraete et al., 2008) and 'arrow of time' (Hall, 2008);

3/ 'Transport' (Zhang et al., 2003) and 'precision' {$\varepsilon = 1/k_c$} (Cvitanovic et al., 2008).

EPS transformation with bond percolation increases EPS fragility. Indeed, the three bond percolation transformations correspond to the EPS changes by (Rosas-Casals and Corominas-Murtra, 2008), which increase its fragility.

EPS transformation causes EPS load change. The load change in bond percolation is defined as non-linear soliton by (Lechtenfeld, 2007). The adiabatic energy of this soliton is found from the rates of optimal steering with two drifts of Zhang and Sastry (Zhang and Sastry, 2002). Security of the grid transformation is determined for the space-time geometry of bond percolation according to (Correa-Borbonet, 2007). Optimal steering is a motion in this geometry. The motion in this space-time geometry is set by the security transformation commutator. This commutator is calculated according to (Belyankov, 2007). Optimal steering is a steering, which is optimal in energy and close to optimal in power, according to (Armitage and Parker, 2007).

Expected synchronous time deviations are obtained from the EPS model as two harmonic oscillators with frequencies $\omega_1$ and $\omega_2$, respectively. EPS frequency as a dipole is Rabi frequency from (Hecht, 2004).

Oracle predicts EPS instabilities from EPS temperature, referred to the Hamiltonian fluctuations of a harmonic oscillator, according to (Bussi and Parrinello, 2008) and (Zhang et al., 2003). From deterministic quantum mechanics of `t Hooft ('t Hooft, 2006) EPS time fluctuations for EPS energy, time and adiabatic energy are found. Hamiltonian fluctuations are obtained from these time fluctuations. Two deviation types from synchronous time are found from the ratios of the various Hamiltonian fluctuations. These deviations correspond to two different EPS instabilities, caused by an external action with frequency $\Omega$ or with frequency $2\Omega/3$. The oracle predicts these two EPS instabilities.

EPS transformation with bond percolation induces current in EPS. First is found the EPS Hamiltonian spectrum according to (Rosu, 2003). EPS current is found from the Parrondo game for EPS Hamiltonian spectrum. Such a game is generated, according to (Toral et al., 2003), from EPS transformation with bond percolation.

EPS expected state is found from the comparison of the EPS transformation with bond percolation dynamics according to (Cramer et al., 2008) and the EPS critical dynamics according to (Budzyński et al., 2007).

Let EPS instabilities are regarded as topological singularities and the transformation with bond percolation - as transport in kicked Harper model. Then the droop is found from the Satija graphics (Satija, 2005) for kicked Harper model.

## 5.2. ENTROPIC STOCHASTIC RESONANCE

Entropy relativistic fluctuations by (Fingerle, 2007) lead to the entropic stochastic resonance in the system 'EPS-Market'. The entropic stochastic resonance in the system 'EPS-Market' is the entropic stochastic resonance for soft matter by (Burada et al., 2008).

The entropic stochastic resonance in the system 'EPS-Market' is a resonance in energy or a resonance in frequency.

### 5.2.1. ENTROPIC STOCHASTIC RESONANCE IN ENERGY

The system 'EPS-Market' is considered as a molecule, composed from four quantum dots, corresponding to the four Regional Dispatching Centers (RDC).

Schematic illustration of this molecule is the same as that of entropic stochastic resonance. Therefore, the entropic stochastic resonance can be represented as such a molecule.

The energy of self-polarization of this molecule is obtained from the EPS droop and from the EPS work to cover the expected load. This energy is derived from the results for such molecules by (Toth et al., 1996) and (Csurgay et al., 2000).

EPS transmission characteristics are the transmission characteristics of the spin filter by (Song et al., 2004). This spin filter is a Fano resonance in open quantum dot. Then the EPS transmission characteristics are probabilities $p_v$ or ($p_r$, $p_v$), found from the limit values of self-polarization energy.

### 5.2.2. ENTROPIC STOCHASTIC RESONANCE OF THE ANGULAR FREQUENCY

The system 'EPS-Market' is considered as decoupled fixed point of the potential load function and of the potential function of the meteorological temperature. The schematic illustration of the decoupled fixed point is the same as that of the entropic stochastic resonance. That's why the entropic stochastic resonance can be represented as such a fixed point.

The potential function here is represented with eight spectral colors. The decoupled fixed point is found with bidirectional associative memory (Blum, 1990). The EPS droop $k_c$ is determined from the expected path for the found decoupled fixed point. The tolerance of this 'EPS-Market' system model is [$k_c$] - tolerance. Almost always this tolerance is 2- tolerance.

The level of coarseness of this decoupled fixed point is found from the one-month memory (Bentz, 1988) for these fixed points. This level of coarseness is another estimation for the 'EPS-Market' system tolerance.

### 5.2.3. FREQUENCY DISTRIBUTION

EPS frequency distribution for a day ahead is determined by present day frequency, present day actual dry-bulb temperature and the forecasted dry-bulb temperature for the next day.

Dry-bulb temperature T corresponds to the length of the white black-body radiation λ. Then (Lavenda, 1991)

$$1/\lambda = T/ (0.393*11.6)$$

Let the EPS frequency f is reciprocal of the length λ and let f-distribution and 1/λ-distribution are Weibull distributions. These assumptions are the result from the entropic stochastic resonance in 'EPS-Market' system. Weibull distribution of EPS frequency for today is recovered (Vapnik et al., 1984) according to the data for this frequency. The 1/λ-distribution for today is recovered (Vapnik et al., 1984) according to the data for the actual dry-bulb temperature.

Distributions of f and 1/λ for today define a mixed distribution (Sum and Oommen, 1995). The mixed distribution of f and 1/λ for tomorrow is found as proportional to the mixed distribution for today. The mixed distribution for tomorrow provides the expected Weibull frequency distribution for tomorrow. The linear relationship between two mixed distributions determines the association between them.

# Chapter III. VERIFICATION

## 1. DAD ACTION VERIFICATION

VERIFICATION 1. DAD separability verification is verification for DAD solving problems and DAD induced change verification.

### 1.1. DAD PROBLEMS SOLVING VERIFICATION

#### 1.1.1. PROOF VERIFICATION

In reality, the man-dispatcher, assisted by DAD, plays (Stefanov, 2010) on cooperation in the game 'Prisoner's dilemma' against another person–dispatcher with DAD. In addition, in reality, the man-dispatcher, assisted by DAD, is (Stefanov, 2010) a security and efficiency arbiter.

This double game of man-dispatcher proceeds in EPS time. Therefore, it is $\varepsilon$-close to the task of finding a maximum clique by (Roughgarden, 2009), $\varepsilon = \max (1/(2k_c), 10(3T_{16}))$. This game has a $2\varepsilon$-Nash equilibrium, which is calculated for polynomial time.

Verification of this Nash equilibrium is to find a correlated equilibrium of the above game and of the approximated equilibrium of the above game. This verification results from the EPS time structure. Correlated equilibrium of the above game is reached by fictitious play by (Christodoulou et al., 2008). Approximated equilibrium of the above game is achieved by a 'congestion' play by (Conitzer, 2009).

The above double game is proved twice - first it is proved quantumly, and then is proved relativistically. This double proof is one round game in between the double evidence and a classical verification.

#### 1.1.2. PREDICTION VERIFICATION

DAD is a triple pendulum by (Valeev and Yumagulov, 2009) with arms, set by EPS time, and masses, set by the EPS expected reliability, in regard to a rare event and a concurrent phenomenon, and EPS expected reliability in regard to cooperative phenomenon. These arms are determined from EPS time according to (Dáger and Zuazua, 2002).

The fluctuations of this triple pendulum demonstrate near-Nash equilibrium and the lower bound of the game value of two-prover game. Herein the lower bound of the game value is obtained according to (Kempe et al., 2007).

Gedanken experiment with EPS for a day ahead codes according to (Dvorský et al., 2009) the EPS viscosity as the prediction depth with this gedanken experiment.

Herein the gedanken experiment is according to (Stefanov, 2010), and the depth prediction is determined by (Moore, 1998). The expected exchanges time $T_{16}$ is 2 to the power ` prediction depth with this gedanken experiment`.

EPS is a large network. Then the specified expected EPS reliability, in regard to the cooperative phenomenon, is found according to (Guitter, 2010) by the droop for a day ahead.

EPS is a network with a complicated security and efficiency optimization. Then the other specified expected EPS reliability, in respect to cooperative phenomenon, is obtained according to (Fabrikant et al., 2002) from the droop and the exchanges time for a day ahead.

The expected synchronous time deviation is a breather or 'soliton – antisoliton', or 'soliton – soliton' a solution of the sine-Gordon equation for a droop and the exchanges time for a day ahead. This solution is obtained according to (Aktosun et al., 2010).

### 1.1.3. BEHAVIOR VERIFICATION

Herein it was decided about the EPS behavior type for a day ahead by Analytical Hierarchical Process (AHP)-protocol.

AHP-protocol glues the load and EPS exchanges in three ways. These three gluing correspond to three different EPS behaviors. Each of these behaviors is characterized by the droop and the exchanges duration in the morning and in the afternoon. These three behaviors are EPS congestion behavior, behavior of market and EPS mutual consideration, behavior of EPS and market mixing.

Decision about the expected type of behavior is taken by auction with AHP-protocol. This decision is obtained using the method by (Goyal et al., 2008).

The type of behavior for a day ahead is defined as an expected deviation of the synchronous time and EPS expected energy deviation, caused by the market.

AHP-protocol is based on EPS coding as droop and time $T_6$. The EPS coding here, is coding according to (Bachok and Vallentin, 2009) in spherical cap, determined by the expected EPS reliability, in respect of a rare event and a concurrent phenomenon and expected EPS reliability, in respect to a cooperative phenomenon.

### 1.2. CHANGE VERIFICATION

The 'EPS-Market' system change is verified as a path, a wave speed and time.

### 1.2.1. PATH VERIFICATION

Expected path of the 'EPS-Market' system is verified by the minimum length of the string, corresponding to the system. It is shown in (Stefanov, 2010) that DAD is an adequate computation of the 'EPS-market' system through diffusion chaos.

### 1.2.2. ADAPTATION VERIFICATION

Dispatcher activities are verified by checking the 'EPS-Market' system adaptation.

'EPS-Market' system adaptation consists of grid organization, when the grid is considered as a complex adaptive system. This follows from the organization results by (Schneider and Somers, 2006). This organization makes the grid similar to urban network by (Zimmermann and Soci, 2004). The similarity of these two networks leads to model similarity with a cellular automata of the particle transport in magnetized plasma (Punzmann and Shats, 2004) and the models with cellular automata of the Schroedinger discrete spectral problem (Bruschi, 2006).

The latest similarity determines the two types of diffusion in magnetized plasma (L-mode и H-mode) by two EPS filtered loads. Herein the two EPS loads are a typical and a non-typical load, low-pass filtered with a filter by (Takalo and Mursula, 2000). The rotation angle between two diffusions is determined according to (Casini and Huerta, 2009).

The urban network is modeled by spin nework, according to (Zimmermann and Soci, 2004). Grid transformation {triangle ↔ star} is presented as a 6j-symbol in spin network. This angle is estimated in regard to rotation angle between two diffusions using the results by (Garoufalidis and van der Veen, 2009) and (Dupuis and Livine, 2009).

Unpredictability as separability (Macleod et al., 2009), is determined by the grid transformation {triangle ↔ star}, according to (Afraimovich and Glebsky, 2003).

The grid edge organization, during management and communication, is with a diameter, determined by unpredictability. This diameter is obtained according to (Dekker, 2007).

In uncertain resources, dispatcher adapts with a constant speed the grid to achieve the edge organization of the grid. This speed is found by (LaValle, 2006) (Lawless et al., 2007) and (Cvitanović et al., 2008).

The grid edge organization, when taking a decision by the dispatcher for grid development, is with the greatest depth, found by (D'Souza et al., 2007). The speed development of the grid is determined by this greatest depth and by the grid development time. Here grid development time is $3T_8$, where $T_8$ is the period of super-harmonic of third order (Elnaggar, 1985) for EPS.

This organization speeds are limited by the fluctuations' speed around thermodynamic equilibrium by (Linden et al., 2009).

Grid organization is non-typical when the two speeds of the edge organization exceed the fluctuations speed around the thermodynamic equilibrium.

Evaluation of the grid organization as a moving target in chaotic or noisy environment is obtained by soliton resonance according to (Zak and Kulikov, 2003).

Organization speed density is presented by (Kontkanen and Myllymäki, 2007) histogram. The first histogram evaluates the grid organization when there is energy turbulence. The second histogram evaluates the grid organization when there is an electricity cost serpent.

Grid organization is evaluated as electrical power transfer time through the results by (Conlon, 2009). The obtained results demonstrate whether this grid is a tunnel and whether this grid is a well for electrical power transfer.

Grid organization is non typical, if the EPS behavior for the previous day has been non typical according to the test by (Vasyechko et al., 2006). This test is performed according to the best load forecast for yesterday and the actual load for yesterday. This test verifies the EPS energy behavior and energy distribution.

EPS self-regeneration is encoded as an association-dissociation in knotted polymer. The polymer intersections number is proportional to the time transfer logarithm to the power of 1.97. The new EPS transfer time is obtained as upper bound of the EPS knotted polymer length. This length is computed (Cantarella et al., 2003) from this polymer intersections number. The knotted EPS energy is found from the universal growth law for knot energy by (Lin and Yang, 2008). According to this law, EPS knotted energy is equal to the EPS knotted polymer intersections number to the power of 0.92.

EPS organization infringer is a kinematic fluctuation of the exchanges that exceeds viscosity from EPS entanglement. This kinematic fluctuation computation is based on the volume of discrete space-time in accordance with (Ahmed and Ridout, 2009) and (Wallden, 2010). This kinematic fluctuation is calculated as well as quantum walk over star network according to (Jafarizadeh and Salimi, 2005) and (Jafarizadeh et al., 2006).

EPS viscosity is obtained in regard to the EPS entanglement entropy in connection to EPS separability or from the EPS stability of character resonances. Here the entanglement entropy is obtained according to (Solodukhin, 2009) and stability of character resonances is obtained according to (Balslev and Venkov, 1999). EPS behavior in regard to volumes is tested through EPS viscosity.

### 1.2.3. PERFECT STATE TRANSFER VERIFICATION

'EPS-Market' system perfect state transfer on a spin chain is realized by complexity 'EPS-Market' system replica. This replica is a replica of the load complexity for the first twelve hours of the twenty-four-hour period and the complexity of the replica exchange for the second twelve hours of the twenty-four-hour period. The perfect transfer is realized by this replica, according to (Di Franco et al., 2009) as an impulse or a step. Complexity load replica gives a spin chain length $N_r$. Complexity exchanges replica gives a spin chain length $N_v$.

'EPS-Market' system perfect state transfer, when there is a cascade defect in it, is realized by a replica from the above type in a spin chain length $N_\Sigma$, $N_\Sigma = N_r + N_v$. Then the number of failed transmission lines during this cascade failure is equal to $N_\Sigma$. The probability of $N_\Sigma$ failed transmission lines is found according to (Dobson, 2007) for branching cascade failure.

### 1.2.4. EXCHANGES VERIFICATION

EPS exchanges are modeled as an electrical transfer through single nanoscale branch point with (Cui et al., 2005) semiconductor tetrapod. Exchange boundary energy is a back-gate voltage function of the tetrapod at boundary differential conductance. This back-gate voltage is calculated from the heat, frequency and output voltage change of EPS.

The heat change is a result from electrochemical potential change, obtained as flag spectra by (Patrão, 2009). The frequency of change is obtained from the probability of $N_\Sigma$ failed transmission lines. The power of change is found by the current of change and the potential of change.

EPS change defines the EPS phase lifetime. Herein, the lifetime is obtained according to (Naud, 2005). This lifetime gives the expected synchronous time deviation, the expected EPS energy deviation, caused by the market, and the expected exchanges time. The expected exchanges time is obtained according to (Naud, 2005b).

### 1.2.5. ENERGY VERIFICATION

EPS exchanges energy is found as extreme value statistics by (Majumdar et al., 2009) for boundary droop $k_{cg}$ and the boundary exchange time $T_{16g}$. This extreme value statistics is determined from the convex hill of $k_{cg}$ random points, $k_{cg} = (3/4)(N_\Sigma)_{max}$, or from the convex hill from $T_{16g}$ random points, $T_{16g} = (N_\Sigma)_{min}$. In the first case the extreme EPS statistics is a mean perimeter of $k_{cg}$ random points in a plane, chosen according to a probabilistic power-law tail distribution. In the second case, the extreme EPS statistics is a mean perimeter of $T_{16g}$ random points, chosen according to a probabilistic Weibull-law tail distribution.

EPS exchanges energy levels, their corresponding electricity cost levels and electricity market temperature, give the synchronous time deviation. This deviation is calculated by the same way as in Chapter II, § 2.

## 2. DAD COMPUTATION

VERIFICATION 2. DAD network verification is DAD imagination verification and DAD predictive program verification.

### 2.1. 'DAD' IMAGINATION TEST

Presenting DAD as a robot and as an intentional agent gives the imagination verification at gedanken experiment by (Stefanov, 2010).

### 2.1.1. DAD AS A ROBOT

DAD checks gedanken association scheme through a network, which is Braided Ribbon Network (BRN). Herein BRN (Hackett and Wan, 2008) is a spin-network, embedded in a manifold.

Perception of BRN provides the association scheme verification through reflection and absorption of light waves by (Fernández-Garsía and Rosas-Ortiz, 2008). Herein the BRN perception is considered as quantum resonance. This resonance is found according to the EPS boundary potential, EPS Bloch angle and the spectral radius of the expected load change. The spectral radius is calculated from failure probabilities ( $p_r$ , $p_v$ ) using the results from (Dekking and Kuijvenhoven, 2008). EPS Bloch angle is calculated by the Nash equilibrium for a game with EPS security. This game with security is a macroscopic quantum game by (Grib and Parfionov, 2008) with costs, reciprocal to failure probabilities. This presentation for failure probabilities results from considering the probabilities as angles by (Feldman and Klain, 2008).

BRN electrochemics provides association scheme verification with asymmetric junction currents by (Kakashvili and Bolech, 2007). BRN electrochemics is considered here as superconduction at welding. Superconduction at welding is obtained by thermalization at welding by (Joshi, 2008), by the thermal current at welding by (Kakashvili and Bolech, 2007) and by superconduction by (Velarde, 2004). Welding thermalization is calculated by the synchronous time deviation, caused by fluctuation uncertainty. Welding superconduction is calculated in regard to the currents, caused by EPS transformation with a bond percolation and in regard to the reverse temperatures of the electricity market.

BRN electrochemics provides as well the second association scheme verification with the directed current of the quasi-adiabatically, run by alternative current, non linear system by (Buttà and Negrini, 2008). The directed current here is calculated according to the expected boundary values of the weather change and from the boundary values of the 'EPS-Market' system energy.

BRN interpretation provides association scheme verification with the path and the heat trace of the expected Nambu mechanical system by (Espindola, 2008) and with the predicted isotropic decay of the turbulence fluctuation by (Eyink and Wray, 1998). The expected Nambu mechanical system is found here from the energy similarity and market behavior similarity. The heat trace of this mechanical system is found for the market thermodynamics, via the results by (Mazzeo and Rowlett, 2009). The predicted isotropic decay of the turbulence fluctuation is obtained from the potential lack of similarity and from this decay time. This potential is calculated according to (Galvez et al., 2004).

The last association scheme verification is by a mechanical tunnel model from (Mouchet, 2008). This is a measurement of the associative scheme, when BRN is a quantum tunnel between two symmetric wells. This quantum tunnel is presented as Klein-Gordon string, where two completely alike oscillators are connected. Beating between two oscillators is considered as a shock in elastic-plastic medium and as a difference between the two 'dervishes' views for the universe. This shock in elastic-plastic medium is obtained from the Sadovskiy's results (Sadovskiy, 1997). The view of the first dervish is a view of a whirling dervish with an angular velocity π/30. The view of the second dervish is a view of a whirling dervish with an angular velocity π/45. Herein the universe and the views towards it are in terms of (Németi et al., 2008).

BRN–verifications validate the imagination at the considered gedanken experiment. These verifications demonstrate that DAD is an imaginary universe. This imaginary universe is Gödel-type rotating universe, which exists in time and is observed by a 'dervish'. This is so, according to visualization of such universes by (Németi et al., 2008).

### 2.1.2. DAD AS AN INTENTIONAL AGENT

Gedanken experiment transforms DAD into an intentional agent. Then DAD is an intention that is universal and is repeating.

The universality of this intentional agent is the universality by (Bizoń and Zenginoğlu, 2008) of the global dynamics for the cubic wave equation.

`ESP-Market` system is ε-complex, according to (Afraimovich and Glebsky, 2003), if it is ε-separable or ε-network. This system is ε-complex, because it is a game by (Parfionov, 2008). That's why `ESP-Market` system is ε-complex as a critical cubic wave by (Bizoń and Zenginoğlu, 2008). `ESP-Market` system is a ε-network because it is with Internet traffic by (Barthélemy et al., 2002). That's why this system is ε-complex as a blowup cubic wave by (Bizoń and Zenginoğlu, 2008). Therefore, the universality of DAD, as an intentional agent, is the universality of global dynamics for the cubic wave equation.

DAD is an EPS dispatcher because it is a GRID-dispatcher, and GRID is EPS analogue. DAD is a GRID-dispatcher that predicts events occurrences for long-term replica optimization in GRID environment.

DAD replica consists of two rules. The first replica concerns EPS complexity for the first 12 hours of the twenty-four-hour period. This replica probability is the EPS estimated reliability with respect to a rare event and a competitive phenomenon. The second replica concerns the EPS complexity for the second 12 hours of the twenty-four-hour period. This replica probability is the EPS estimated reliability with respect to a cooperative phenomenon.

This DAD replica is a GRID-dispatcher replica by (Ciglan et al., 2005).

DAD replica stabilizes the estimated electricity price at fixed EPS droop. Herein the price stabilization is realized by dividing the twenty-four-hour period into three periods. These three periods are: EPS noon maximum load, EPS afternoon load drop and EPS evening maximum load. Thus, the estimated price stabilization can be found from the results by (Rozdilsky et al., 2004) for Lotka-Volterra replica. The stable price shows that 'EPS-Market' system is efficient.

EPS security optimization with this replica is likely to be an acceptable search in two regions. The number of probably acceptable heuristics for this search is the efficient exchanges time. This number is defined according to (Ernandes and Gori, 2004).  The two regions, where the search is performed, are determined by the quantum tunnel width. This width is found as a link width according to (Lackenby, 2008).

Thus DAD as a robot and an intentional agent assists the man-dispatcher with EPS security and efficiency predicting analysis. Therefore, DAD gedanken experiment verification is successful.

## 2.2.  'DAD' PREDICTING PROGRAM TEST

Expected synchronization and expected 'EPS-Market' steady state, considered as 3ε-network, provide the expected S/U–system state and the expected synchronous time deviation $dt_s$ for such a state. Herein S/U state (Sleep/Update - state) is an EPS-Market system state when there are incessant changes in the system.

This model system 'EPS-Market' is derived from the system angular frequencies. The system angular frequencies are transformed into powers by AHP (Bozóki and Lewis, 2005). The system angular frequencies are the two main frequencies, the Rabi frequency and frequency 1.291 by (Bachelard et al., 2008).

This AHP- leading causes 'EPS-Market' system behavior, which is developed in closed time like geodesics by (Grøn and Johannesen, 2010). Indeed, (Smolyakov, 2009) AHP-leading provides extreme system behavior, when the system does not change.

The gain and the system phase are those for plasma with the same frequencies. They can be found using (Fukagata an al., 2008) at efficiency of the third harmonic, which is 6.93949 and 1.0/0.124504 (8.03187), where the third harmonic efficiency is that from (Ondarza, 2005).

Let EPS is a network with a star structure. Let the exchanges in this network are fast solitons by (Adami et al., 2010). Then, when there is AHP- leading, EPS travels path r for a time t and the market travels path S for a time T. Here the two paths and the two times are determined according to (Adami et al., 2010).

Let 'EPS-Market' system is a hyperbolic network by (Baryshnikov and Tucci, 2010). Then the 'EPS-Market' system state, at expected AHP-leading, is obtained from the expected action level on the system as the state of quantum spin chain having a local interaction between three Ising spins and longitudinal and transverse magnetic fields. This level is set by the number of actions on the system at AHP-market leading with a path S for a time T. Here the action level is determined according to (Rokach et al., 2008), and the quantum spin chain state is according to (McCabe et al., 2010).

The EPS droop is found by the phase transition in Ising magnetic by (Kozlov, 2008).

Accepting probability and success probability for AHP-market leading, can be found from considering the phase transition in Ising magnetic as quantum program branch by (Sasaki, 2002) и (Ablayev and Vasiliev, 2008). The verification for expectations of acceptance and success of the AHP-market leading is achieved according to the results for quantum and stochastic branching programs of bound width by (Ablayev et al., 2002). These two probabilities are represented as a loop by (Aalok, 2010), (Po et al., 2007) and (Rennen et al., 2009).

The loop connects 'DAD as a WATCH' with 'DAD as an ASSISTANT'. DAD as a WATCH sleeps with one eye open. Then DAD takes a decision in partial ignorance, according to (Gigerenzer, 2007). DAD as ASSISTANT updates the droop and exchanges so, that the phase is constant at fixed EPS energy. Then DAD successfully leads the system, according to (Gigerenzer, 2007). The leading with a constant phase is AHP-leading, according to (Smolyakov, 2009). Because the updating is a kind of a revision, then DAD is an ASSISTANT in sense of (Arvo, 1999).

Let the 'EPS-market' system rate of change at AHP-leading is set by the weather rate of change. This change is described by perturbed soliton solution of the 'sin-Gordon' equation by (Popov, 2009).

Energy and energy perturbation of this system change are the energy and energy perturbation of perturbed soliton magnetic flux. The temperature of this magnetic flux is the traveling wave temperature before the piston by (Volosevich and Levanov, 1997).

EPS electromagnetic field temperature, at AHP-leading, is provided by heat kernel transformation of the Heisenberg group by (Krötz et al., 2004).

Let the 'EPS-Market' system state change is with Langevin dynamics for Markovian state model. The probability for this change is determined by (Singhal et al., 2004) at the transition from AHP-leading for three angular frequencies to AHP-leading for four angular frequencies. It is accepted that at this AHP-leading change, the system temperature is changed from the magnetic flux upper temperature to the electromagnetic field upper temperature. The acquired probability p is the probability of the system S/U–state.

The expected S/U–state for today is obtained from mixing the expected U-state for today with S/U–state for today with the probabilities (1-p) and p, respectively. In this way the expected 'EPS-Market' system state is specified according to (DeVille and Mitra, 2009).

'EPS-Market' system, considered as 3ε-network, is decomposed into an infinite line and 3-regular tree. The growth of this network is the synchronous time deviation $dt_s$. The above system decomposition provides the expected $dt_s$ boundary values as heat kernel values of 'homesick random walk' by (Brasseur et al., 2009) and (Lownes et al., 2007).

Considering the system as a 3ε-network, makes it ε-optimal, according to (Sorin et al., 2010). The norming of the $dt_s$ expected boundary values is obtained from above.

S/U–state of 'EPS-Market' system is perceived as a state from exceptional point proximity. In this case perception is through stochastic resonance by (Jian-Hua and Xian-Bin, 2010), where the state of proximity of exceptional point is determined according to (Heiss, 2010). Therefore DAD predicts a decay state and EPS overload state.

The subharmonic oscillations of 'EPS-Market' system are classified according to (Kiselman, 2004) by period and amplitude, where the amplitude is a droop and signal-to-noise ratio.

The electricity cost is obtained as a transfer in the network for a time T. Herein the transfer is a solution for time-fractional Burgers equation by (Wu, 2010). The fractional degree of this equation is departure from Gaussian noise (Xie et al., 2009), deviation which corresponds to stochastic resonance in the system.

Synchronous time deviation $dt_s$ presents the entropy in the system at it normal operation. This entropy is found according to (Adivar, 2010) from the fractional degree and the normalized exchange time.

S/U–state of 'EPS-Market' system is a genuine non-equilibrium state under feedback control. Then the thermodynamics by (Abreu and Seifert, 2011) provides the expected extra dispatching information, which DAD carries as a gedanken experiment. This expected extra dispatcher information refers to anticipated rare events, competitive events and cooperative phenomena. The expected EPS reliability in regard to a rare event and a competitive phenomenon, and the EPS expected reliability, in regard to a cooperative phenomenon, are obtained from the expected dispatching extra information when considered by (Ruseckas et al., 2011) the expected rare events, competitive and cooperative phenomena.

### 2.2.1. SYSTEM COHERENCE

'EPS-Market' system is considered as a quantum network, connected with the $V_{24}$ group to test the expected synchronization and the expected 'EPS-Market' system steady state, considered as a 3ε-network.

Pure state transfer in this quantum system is determined by the elasticity and plasticity of the network. This transport is lossless transfer in the 'EPS-Market' system. Then the lossless transport is obtained using the phase of the 'EPS-Market' system via the results for pure state transport of (Jafarizadeh et al., 2010) and (Brezis and Peletier, 2006).

The acceptance and the pure transport state success depend on the knot and curtailing links of this quantum network. Therefore, the probability of pure transport state acceptance is found for magnetic knot by (Candelaresi et al., 2010) and the likelihood for pure transport state success is determined for graphical game by (Dilkina et al., 2007). These two probabilities are presented as a loop by (Aalok, 2010), (Po et al., 2007) and (Rennen et al., 2009).

'EPS-Market' system, as a quantum network with a pure transport state, is a Wigner quantum network by (Regniers and Van der Jeugt, 2009). Then it is a system with coupled harmonic oscillators with constant interaction. The phase turbulence (Glyzin et al., 2010) in it gives the exchange 'EPS-market' system time. The flow in it (Glyzin et al., 2010) gives the droop, the electricity cost and the signal / noise ratio of the 'EPS-Market' system. Its operation as Otto quantum cycle (Quan et al., 2006) gives another signal / noise ratio of the 'EPS-Market' system. Its operation as a quantum forced oscillator (Campisi, 2008) provides the second electricity cost.

Thus is restored the 'EPS-Market' network coherent state as a coherent state of the Maxwell field by (Finkelstein, 2010). This 'EPS-Market' network coherent state is safe and effective state.

The above loop comparison, with the predictive program loop, shows whether the 'EPS-Market' network is a coherent system.

### 2.2.2. SYSTEM SELF-HEALING

Structural self-organization of the 'EPS-Market' system (self-healing) is tested via minimization according to (Finster and Schieferender, 2010), based on causal variational principle. This minimization generates spontaneous formed structure according to (Finster and Schiefereneder, 2010).This spontaneous formed structure is perceived through the EPS characters resonance by (Balslev and Venkov, 1999) and is interpreted as a quantum catastrophe by (Emary, 2005).

Deviation from the synchronous time for self-organization is obtained from the critical parameters of the 2D-catastrophe by (Emary, 2005). When the structural self-organization occurs in the 'EPS-Market' system, this deviation is less than the deviation 'EPS-Market' system synchronous time, due to its nonintegrability. The 'EPS-Market' system synchronous time deviation, due to its nonintegrability, is found according to (Corrigan and Zambon, 2010) as a delay due to integrable defect in the system.

'EPS-Market' system entropy, caused by nonergodicity and nonintegrability of this spontaneous formed structure system, is determined as entropy by (Finster and Schiefereneder, 2010) and (Emary, 2005).

DAD perceives the structural self-organization of the 'EPS-market' system through gedanken experiment that is 2-cyclic design at analytical hierarchical process of decision making by (Miyake et al., 2003). The thermal resonance by (Kirsanov, 2009) between a system and environment causes this perception. Entropy of this perception is determined by the standard error (Miyake et al., 2003) of 2-cyclic design in analytical hierarchical process of decision making, which single cycle is with a length of 27.

DAD action, at 'EPS-Market' system structural self-organization, is the minimum action on the sphere by (Finster and Schiefereneder, 2010).

DAD leads structurally self-organized 'EPS-Market' system securely and efficiently, when the entropy of its action is close to the entropy of its perception and to the entropy, caused by nonergodicity and nonintegrability of the spontaneous formed structure of the 'EPS-Market' system. Herein the entropies are close as entropy approximations by (Finster and Schiefereneder, 2010).

'EPS-Market' system structural self-organization leads to predictability for a day ahead, according to (Breitung and Candelon, 2006).

# Chapter  IV.  VALIDATION

1.  DAD LOGIC

VALIDATION.  EPS security and efficiency quantum/relativistic computation for a day ahead is valid, when it predicts a threat through the expertise of the EPS world and when it evades the threat via an interactive game for resources.

1.1. DOUBLE PROOF OF DAD

DAD obtains the expected EPS quantum state through probabilistic three-query. Indeed, DAD is an interactive proof of type 'oracle', which is performed by two-prover one-round between them. Then DAD predicts by the above technique according to the results for an interactive proof by (Ito, 2008). Because DAD is such a system of proof, then DAD is a reliable partner/rival and provides the man-dispatcher with the advantage to be an arbiter.

The three questions are for visualization (synchronization), interpretation (stabilization) and modeling (essence) of the expected system 'EPS-Market'.

These questions are in two options. The first option is for 2-tolerance of visualization, interpretation and modeling. This question is put by the ASSISTANT at entropic stochastic resonance, which determines the electricity transport and the market landscape. The second option is for a predicting market at visualization, interpretation and modeling. This question is formulated by the WATCH at virtual thermalization, determining surprises in EPS and the network connections percolation.

The ASSISTANT is tolerant, because the changes are ambiguous. So it can ask for 2- tolerance.

The WATCH is perspicacious. Therefore, it can ask about a predictive market, according to the    Ramsey-type results on random graphs by (Berarducci et al., 2008) and the results for predictive market by (Stefanov, 2007).

The language of the ASSISTANT and the WATCH is a language of unitary gate by (Fiurášek, 2003) and two-strand system of DNA type by (Aerts and Czachor, 2005). This language is a language of Wilson loop for D5-brane by (Faraggi and Zayas, 2011).

DAD distinguishes the expected threats for EPS and raises the alarm. DAD distinguishes the expected threats according to the expected EPS states and the expected market states. The expected EPS states are normal, restorative and emergency. Market is with the same expected states, according to (Sobajic and Douglas, 2004).

The alarm raised by DAD (Stefanov, 2010b), is with the following levels: low, guarded, elevated, high, and severe. These alarm levels coincide with the levels from 'Threat Alert System' (http://www.esisac.com).

These levels are derived from coding of the expected states with two bits and from the results of (Ito et al., 2008). At this coding the cooperative phenomenon is considered as unknowable, and the rare and concurrent phenomenon as representable via a diagram. Then there are the following thresholds:

1) EPS reliability with respect to a cooperative phenomenon $p_v$ is less than 0.36. Then the cooperative phenomenon is unknowable. The unknowability (Appleby, 2003) is with $\sigma=0.12$ and herein the boundary value is $3\sigma=0.36$.

2) EPS reliability, with respect to a rare event and concurrent phenomenon $p_r$ is less than 0.85. Then the rare event and the concurrent phenomenon are presented as diagrams, but not as histograms. Histogram is derived from (Lejeune, 2007) with a probability greater than $1-18^{-2/3} = 1-0.145 \approx 0.85$. Then, a diagram is obtained with a probability less than 0.85.

DAD quantum logic is superbraid quantum logic by (Yepez, 2009). EPS is a ribbon of this superbraid, and the ASSISTANT and the WATCH are its two qubits.

DAD verification as two-prover one round interactive proof system is according to (Chemero and Eck, 1999).

The first DAD proving device is presented as gedanken experiment, generated by adaptive resonance (Bussi and Parrinello, 2008) with the environment. The second DAD proving device is presented as emulator, generated by entropic resonance ('t Hooft, 2006) with the environment.

This double proof device is a physical world with two time dimensions by (Foster and Müller, 2010). Geometric mechanics of thid world is geometric mechanics on a product Heisenberg groups by (Chang et al., 2009). The choice in this world is rational according to (Jones, 1999).

1.2. DAD'S WORLD

DAD's world is a debugger for 'EPS-Market' system that stops when there is a defect in it for a day-ahead. This world is almost certainly chaotic, according to the results from a debugger by (Mondal and Ghosh, 2011).

DAD's world logic is prediction/evasion logic of Brendle and Shelah (Brendle, 1995), (Brendle and Shelah, 1996, 2003a, 2003b). Prediction/evasion leads to double diagonalization of the quantum/relativistic computation.

At the quantum computation, the angular velocity and energy permanence $<\omega_{t+1}=\omega_t>\&<E_{t+1}=E_t>$ are used for double diagonalization.

At the relativistic computation the equation $<s=t=(kc)½>$ serves for double diagonalization. In this equation s is a path, t is a time, $k_c$ is an EPS droop. In Planck scale physics by (Hossenfelder, 2006) is valid the equation $<(s)^{½}/m_p \sim (t)^{½}/m_p \sim 1>$, where s is a path, t is a time and $m_p$ is Planck mass. The equation from the relativistic computation is similar to this equation when considering the EPS droop as a mass to the power of four.

Two diagonalization couples determine DAD's world complexity as synchronization and steady state by (Stefanov, 1992, 1994). Then DAD's world is a world with two schemes by (Bozóki and Lewis, 2005) and two scales by (Po et al., 2007), and the successful EPS leading in DAD's world is achieved through analytical hierarchical process for taking a decision about the behavior type by (Goyal et al., 2008) and for the causal action by (Miyake et al., 2003). Successful leading here means secure and efficient leading. This leading is leading according to a calendar. EPS leading according to a calendar is obtained in (Stefanov, 2002).

EPS calendar is heuristics to solve the game by (Roughgarden, 2009) for obtaining the maximum clique. EPS calendar leads to ε-Nash equilibrium of this game, which is calculated for a polynomial time. In this case $ε = \max (1/k_c, 20/(3T_{16}))$, $k_c$ is an EPS droop , and $T_{16}$ is exchanges time.

EPS calendar sets a successful EPS leading that is 1/ε-tolerant towards ε-separability of 'EPS-Market' system and with 1/ε-branching of the 'EPS-Market' system ε-net. This successful leading is ε-complex according to the results by (Afraimovich and Glebsky, 2003).

Successful leading is derived by integrable quantum computation and one-way quantum computation. Indeed, DAD is a superbraid (§2.1) and, according to (Zhang, 2011), integrable quantum computation. DAD is obtained via virtual thermalization, as well, and according to (Markham et al., 2010), is one-way quantum computation.

The droop (1/ε) harmonizes (Dijkstra, 2010) energy and time when leading is successful. This harmonization is observed as embedding brane in flat two-time space by (Andrianopoli et al., 1999).

EPS calendar is similar to Kananthai generator, created by two Riesz kernels (Satsanit and Kananthai, 2010). EPS state, at its successful leading with this calendar, is 11D statistical state of the electromagnetic field by (Brody and Hughston, 2009). This similarity suggests that the calendar of the EPS leads to a game of social interactions with local and global externalities by (Le Breton and Weber, 2009).

Dad's world is tested through DAD non-monotonic refinement by (Marincic et al., 2007). At this refinement, DAD is built by the two diagonalization pairs for 'EPS-Market' system that is knowable with the gedanken experiment by (Stefanov, 2010a). This DAD design is a game of knowing the world by DAD (Stefanov, in press). This game of knowing has (Tian, 2009) Nash equilibrium because of 'EPS–market' system diagonalizations.

## 2. DAD TOY-MODEL

PARTIAL VALIDATION 1. Quantum/relativistic computation of EPS security and efficiency for a day ahead is partially valid when it is recorded in PCF-style (Programming language for Computable Functions-style computation).

Dad's world, considered as a string, gives DAD toy-model. Then the string is presented as topology and landscape. Free energy string field is investigated by this presentation, at set left boundary distribution of free energy. The free energy string is determined according to (Grassi, A., 2002). This field is revealed from the dialogue between Player and Opponent with an innocent strategy game by (Harmer, R. et al., 2007). In this case the strategy dialogue is innocent because it is assumed that the topology and the landscape characterize ARIA, and that they are completely visible. ARIA topology is seen as a cell rectangular network and the landscape as – a spectral amplitude rectangular network. The naive strategy action is amplitude switching, corresponding to the spectral amplitude at the same landscape point. The spectral amplitude is found using (Hordijk and Stadler, 1998) from the landscape correlations, calculated using (Kedem, 1994). Switching amplitude is obtained via harmonic linearization according to (Solodovnikov, 1969). This landscape is similar to the free energy landscape of a large-Q Potts model by (Bauer et al., 2009).

The restored free energy string field gives the right boundary distribution of the free energy, corresponding to the set left boundary distribution. Multiple pairs of boundary distributions represent the string gauge. These pairs are ARIA creatures.

From this gauge, calibrated according to M.Barany ( http://www.math.umn.edu/~reiner/ ),  is obtained the string knot Tutte-polynomial that is a knot of the above dialogue. ARIA extension is obtained using (Jaffe et al., 2009) as string stretching $\gamma$. This stretching is obtained by the number of 0-1-2 increasing trees upon 23 vertices $T_{23}(1,-1)$ and from the number of acyclic orientations with unique source upon 22 vertices $T_{22}(1,0)$. It is calculated according to (Merino, 2008) from the critical configuration of Potts model

$$\gamma = (1 + (\log_2(d_G))^2)^{\frac{1}{2}}$$
$$d_G = \log_2(T_{21}(2,-1) - T_{21}(2,0))$$

String extension gives the naïve strategy horizon.

String superconductance temperature in relative units, is

$$T = 1.261060863/\exp(2 - \gamma/3)$$

The superconductance corresponds to ARIA percolation, in correspondence with (Gliozzi, 2006).

This temperature is the temperature of the above dialogue. It resembles the temperature in the game 'Cold, cold, warm…'.

This dialogue is 2-cyclic design in AHP by (Miyake et al., 2003). Indeed, the random experiment with a toy-model provides a minimal temperature Ta=0.69394, and the minimal temperature at 2-cyclic design Q1+Q3 for forty items is Tb=0.6932, according to (Miyake et al., 2003).

Also, this dialogue is a prediction/evasion at mutual information by (Foster and Grassberger, 2010). This is so, because the minimal temperature for prediction/evasion at mutual information for random sequence with a length equals to 11, is Tc=0.6975, according to (Foster and Grassberger, 2010).

The naïve dialogue strategy is set by a clockwork demon by (Morikuni and Tasaki, 2010), that restores the free energy string field. This demon is an EPS toy-calendar.

DAD's toy-model is calculated in ARIA PCF-style (Programming language for Computable Functions-style computation) (Hyland and Ong, 1993). In this calculation ARIA uncertainty is considered according to (Flegontov, A.V. et al., 2008). ARIA non-factors are modeled as a string gauge, ARIA's dispatcher is substituted by cell rectangular network (Harmer, 2007), ARIA non-linear dynamics is modeled with landscape correlations, and the statistical description is presented by Tutte-polynomials.

DAD toy-model predicts EPS load for a day ahead (Stefanov, 1997).

3. DAD'S NEAR PRECISE CALENDAR

PARTIAL VALIDATION 2. Quantum/relativistic computation of EPS security and efficiency for a day is partially valid, when the coarsening 'EPS-Market' system acts bounded rationally.

Near precise EPS calendar globalizes bounded rational DAD's world and makes coarsening the DAD's world network model. In this way is obtained nearly successful EPS leading.

DAD's world is bounded rational world, because it is not completely predictable.

Completely rational world is globalized bounded rational world. DAD bounded rational world globalization is achieved by the calendar via load developing, completely determined by EPS droop. DAD's world at such globalization is in transition state by (Liverts and Barnea, 2011). This globalization is obtained via convergent decentralized design by (Gurnani and Lewis, 2008).

Because initially DAD's world is bounded rational, DAD obtains Pareto-optimal security and efficiency for a day ahead. When DAD's world is developed in completely rational, DAD obtains Nash-optimal security and efficiency, which give way Pareto-optimal values.

This globalization is presented as a crystal growth/melting by (Li et al., 2004). From this growth/melting is obtained the ratio of energy to entropy in completely rational world. As well, from this growth/melting is determined the expected half duration for a day ahead. Therefore this growth/melting is considered as energy transport in a hierarchical lattice by (Pal et al., 2011).

Work at this globalization is obtained according to (Vaikuntanathan and Jarzynski, 2011). Completely rational DAD's world temperature is found from this work.

Globalization is a fair cost sharing game and congestion game. DAD selfishness levels at this globalization are determined according to (Apt and Schäfer, 2011).

'EPS-Market' system is with three states, determined from DAD selfishness levels.

The essence of this globalization is a quantum morphogenesis, according to (Aerts et al., 2003), of the whole 'EPS-Market' system. Kinetic and potential energy of this whole system are defined according to (Sun et al., 2011), (Shmatkov, 2011) and (Bachelard et al., 2008). Probability of being stable the whole system is determined by (Aerts et al., 2003). Admissible stability margin of the whole system is obtained according to (Sun et al., 2011).

DAD's world network model is coarsened by lattice integration according to (Hietarinta and Viallet, 2011). The coarser world is also DAD's world in sense of (Dijkstra, 2010).

This globalization leads to a new EPS design. Considering globalization as design by (Joseph and Hung, 2008) gives a new EPS droop. Considering globalization as a symmetry group $S_3$ by (Kornyak, 2011) provides a new EPS phase angle and irreversibility assessment of the globalization as irreversibility according to (Garmon et al., 2011).

This coarsened 'EPS-Market' system is a quantum system with two levels. This system is controlled by EPS reliability in regard to a rare event and a concurrent phenomenon $p_r$ and EPS reliability in regard to a cooperative phenomenon $p_v$, presented as sine-waveform controls by (Zhang et al., 2010) for a two–level quantum system. Then $p_r$ is explained by the EPS morning leading on the estimated load, and $p_v$ is explained with the EPS estimated afternoon synchronization. The coarsening 'EPS-Market' system entropy for a day ahead is obtained according to (Budzyński et al., 2007). This entropy assesses the coarsening according to (Shany and Zamir, 2011).

**CONCLUSION**

An algorithm for quantum/relativistic security and efficiency computation for Electrical Power System (EPS) is built in this work. This algorithm consists of following steps:

1/ ENERGY RENORMALIZATION

Virtual thermalization of the 'Elastic/Plastic' model of the 'EPS – Market' system: a/ provides the EPS critical load; b/ outlines the EPS fractal landscape.

2/ TIME RENORMALIZATION

Virtual thermalization of the 'Energy/Cost' model of the 'EPS – Market' system: a/ gives the critical exchanges in EPS; b/ founds the minimal electricity leveled cost and the electricity market price.

3/ FLOWCHART

Homogenization of the market leads to a steady state market. Electricity market surprises synchronize EPS with the market. Random Matrix Theory (RMT) - communication of market participants synchronizes the electricity market with EPS. Electricity flow, determined by the market, sets EPS in a steady state.

This flowchart causes entropic stochastic resonance in the system 'EPS-Market' and makes EPS bond percolation.

4/ VERIFICATION

DAD separability verification is verification for DAD solving problems and DAD induced change verification.

DAD network verification is DAD imagination verification and DAD predictive program verification.

5/ VALIDATION

EPS security and efficiency quantum/relativistic computation for a day ahead is valid, when it predicts a threat through the expertise of the EPS world and when it evades the threat via an interactive game for resources.

Quantum/relativistic computation of EPS security and efficiency for a day ahead is partially valid when it is recorded in Programming language for Computable Functions (PCF)-style and when the coarsed 'EPS-Market' system acts bounded rationally.